\documentclass{elsart}
\usepackage{amsfonts}\usepackage{amsbsy}
\usepackage{mathrsfs}
\usepackage{graphicx}
\def\cal{\mathcal}

\newcommand{\lsim}{\stackrel{\scriptstyle <}{\phantom{}_{\sim}}}
\newcommand{\gsim}{\stackrel{\scriptstyle >}{\phantom{}_{\sim}}}

\makeatletter
\def\ps@copyright{\let\@mkboth\@gobbletwo
  \def\@oddhead{}%
  \let\@evenhead\@oddhead
  \let\@evenfoot\@oddfoot
}
\begin{document}
\begin{frontmatter}



\title{Fluctuations
of the Color-superconducting Order Parameter in Heated and Dense
Quark Matter}

\author{D. N.~Voskresensky$^{1,2}$}
\maketitle

\noindent                        
$^{1}${\it\small Gesellschaft f\"ur Schwerionenforschung mbH, Planckstr. 1,
64291 Darmstadt, Germany}
\\
$^{2}${\it\small Moscow Institute for Physics and Engineering, 
Kashirskoe sh. 31, Moscow 115409, Russia}



\begin{abstract}

Fluctuations of the color superconducting order parameter in dense
quark matter at finite temperatures are investigated in terms
of the phenomenological Ginzburg - Landau approach. Our estimates
show that fluctuations of the di-quark gap may strongly affect
some of thermodynamic quantities even far below and above the
critical temperature.  If the critical temperature of the di-quark
phase transition were rather high one could expect a manifestation
of fluctuations of the di-quark gap in the course of heavy ion
collisions.

\end{abstract}
\end{frontmatter}

PACS number(s): 11.15.Ex, 12.38.Aw, 12.38.Mh, 25.75.-q,
74.40.+k

keywords: color superconductivity, order parameter, temperature, fluctuations


The quark-quark interaction in the color antitriplet channel is
attractive driving the pairing, cf. \cite{bl}. The problem has
been re-investigated in a series of papers following Refs. \cite
{arw98,r+98}, see review \cite{RW} and Refs therein. The
attraction comes from the one-gluon exchange, or from a
nonperturbative 4-point interaction motivated by instantons
\cite{dfl} or nonperturbative gluon propagators \cite{br}. The
zero-temperature pairing gap $\Delta$ was predicted to be $\sim
(20\div 200)$~MeV for the quark
chemical potentials $\mu_q \sim (350\div 500)$~MeV.
The critical temperature can be estimated as $T_c \simeq 0.56 \Delta$, 
as in the
standard BCS theory, cf. \cite{RW}.

One expects the di-quark condensate to dominate the physics at
densities beyond the deconfinement/chiral restoration transition
and below the critical temperature. Various phases are possible.
E.g., the so called 2-color superconductivity (2SC) phase allows
for unpaired quarks of one color. There may also exist a
color-flavor locked (CFL) phase \cite{arw99} for not too large
strange quark masses \cite{abr99}, where the color
superconductivity (CSC) is complete in the sense that the di-quark
condensation produces a gap for quarks of all three colors and
flavors. The value of the gap is of the same order of magnitude,
as that in the two-flavor case.

The high-density phases of QCD at low temperatures may exist in
interiors of most massive neutron stars affecting the cooling,
rotation and magnetic field, cf. \cite{bgv,bss,IB}. It is also
possible to ask whether CSC is relevant for the terrestrial experiments?
To produce a color superconducting matter  in the laboratory one
would need to cook a dense and not too hot baryon enriched matter.
The nuclear matter prepared in heavy ion collisions at SIS100,
AGS, SPS or RHIC in all the cases has presumably not sufficiently high
baryon density and, on the other hand, the matter prepared at SPS
or RHIC  is feasibly  too hot in order to expect a manifestation
of the CSC.
The most relevant is probably the GSI ``Compressed baryon matter'' 
heavy ion collision
facility which may cook enough dense and not too hot quark-gluon
matter. The baryon densities up to ten normal nuclear matter
density ($\lsim 10\rho_0$) at temperatures $T\lsim 170$~MeV are
expected to be reached at an initial collision stage. It is
supposed that the system is in the quark-gluon plasma state at
such conditions. Then the system expands and cools down. In this
process the temperature decreases  up to $T\sim 140$~MeV at still
rather high baryon density at an intermediate collision stage.
Although the temperatures at relevant densities are most likely
larger than the critical temperature $T_c$ of CSC, one may rise
the question on a possible manifestation of the precursor
phenomena of the CSC phase transition, if $T_c$ is rather high
($T_c \gsim (50\div 70)$~MeV). Recently, ref. \cite{KKKN02}
considered such a possibility within the Nambu-Jona-Lasinio  model
and demonstrated that the fluctuating pair field  results in a
prominent peak of the spectral function, which survives in the
temperature interval $|T-T_c|\lsim 0.2T_c$. Besides, it is
interesting to investigate the role of the order parameter
fluctuations for $T<T_c$, if $T_c$ is rather low ($T_c \lsim (30\div
50)$~MeV), that may affect the neutrino radiation of the most
massive hot neutron stars at an initial stage of their evolution.

In this Letter we will study precursor phenomena of the CSC
in the framework of the phenomenological Ginzburg - Landau
approach. To be specific, we will consider condensates with 
total angular momentum $J=0$ that are antisymmetric in color and
flavor. Such pairing states can occur in the weak coupling limit
because one-gluon exchange is attractive in the color anti-triplet
channel.

The  Fourier component of  the density of the thermodynamic
potential (thermodynamic potential per unit volume $V$) in the
superconducting quark matter with the di-quark pairing  can be
written in the following form \cite{IB,GR}, cf. \cite{bl,bss},
\begin{eqnarray}  \label{fe}
\widetilde{\Omega}&=&\widetilde{\Omega}_n + \sum_{\alpha
,i}(-c_0|\partial_{\tau} d_{\alpha}^{i}|^2 +c |\nabla
d_{\alpha}^{i}|^2) +a D + \frac{b}{2} D^2 ~,\,\,\,\,
\\
D &=&\sum_{\alpha ,i}|d_{\alpha}^{i} |^2 ,\,\,\, \gamma =
\frac{1}{D^2}\sum_{\alpha , \beta ,i} |d_{\alpha}^{i, *}\cdot
d_{\beta}^i |^2 ,\,\,\, b=b_1 +\gamma b_2\nonumber \,.
\end{eqnarray}
The Greek indices $\alpha ,\beta =\{R,B,G\}$ count colors, the
Latin indices $i =\{u,d,s\}$ count flavors. The expansion is presented
up to the fourth order in the di-quark field operators (related to the gap)  
assuming the
second order phase transition, although at zero temperature 
the transition might be of the first order, cf. \cite{PR}.
$\widetilde{\Omega}_n$ is the
density of the thermodynamic potential of the normal state. 
The
order parameter squared is $D=|\vec{d}_{\rm IS}|^2 =|\vec{d}_R|^2
+|\vec{d}_G|^2 +|\vec{d}_B|^2$, $\vec{d}_R \| \vec{d}_G \|
\vec{d}_B$ for the isoscalar phase (IS), and $D=3|\vec{d}_{\rm
CFL}|^2$, $|\vec{d}_R|^2 =|\vec{d}_G|^2 =|\vec{d}_B|^2
=|\vec{d}_{\rm CFL}|^2$, \,\,$\vec{d}_R^* \cdot \vec{d}_G
=\vec{d}_G^* \cdot \vec{d}_B =\vec{d}_B^* \cdot \vec{d}_R =0$  for
the CFL phase, $\vec{d}_{\alpha} =\{d_{\alpha}^u , d_{\alpha}^d ,
d_{\alpha}^s\}$. The so called 2SC phase is the IS phase with
unpaired quarks of one color (to say $R$).  
The interaction of the di-quark field with fluctuative gluon fields
is usually introduced in the standard way
through the corresponding gauge-shifted derivatives. Recent paper
\cite{GR03} demonstrated that the gluon fluctuations contribute essentially to
thermodynamic quantities only at
temperatures in a 
narrow vicinity of the critical temperature, $\Delta T/T_c  \lsim 0.1$
for relevant values of parameters. In this small temperature interval, 
$\Delta T $, gluon field fluctuations change the nature of the phase
transition (from the second to the first order).
However the value $\Delta T$ is much less than the
temperature region where fluctuations of the di-quark order parameter
might be  important, as we will show below. 
Thereby in the
latter discussion
we suppress the gluon fields as well as the discussion of any
peculiarities of this narrow temperature region near $T_c$.

Near the critical
point coefficients of (\ref{fe}) can be expanded in
$t=(T-T_c)/T_c$. In the weak coupling limit they render:
\begin{eqnarray}  \label{fe-coef}
a&=&a_0 \mbox{ln}(T/T_c )\simeq a_0 t,\,\,\, a_0 =
\frac{2\mu_q^2}{\pi^2},\,\,\,\, b_1 =b_2 =b/(1+\gamma )
=\frac{7\zeta(3)\mu_q^2}{8\pi^4
T_c^2},\,\,\,\,\nonumber \\
 c&=&\frac{b}{3(1+\gamma)},\,\,\, c_0
=3c.
\end{eqnarray}
For the classical (mean) fields $\gamma =1$ in the IS phase and
$\gamma =1/3$ in the CFL phase, cf. \cite{IB}, $\mu_q =\mu_B /3$,
$\mu_B$ is the baryon chemical potential (the contribution of the
strange quark mass is neglected), $\zeta (3)=1.202...$ The
critical temperature $T_c$ is the same for the IS and  the CFL
phases for $m_s \rightarrow 0$. For $m_s \neq 0$, $T_c$ would depend on
$m_s$ that would result in a smaller  $T_c$
for the CFL phase than for the IS phase.

The  value of the order parameter follows by solving
the  equation of motion for the field operators, $\delta \widetilde{\Omega}
/\delta d_{\alpha}^i =0:$
\begin{eqnarray}\label{meq}
-c_0 \partial^2_{\tau} d^i_{\alpha} +  c\Delta d^i_{\alpha} -a
d_{\alpha}^i-b D d_{\alpha}^i =0.
\end{eqnarray}

The  stationary, spatially-homogeneous mean field solution of (\ref{meq})
(without taking
into  account of fluctuations)   is
\begin{eqnarray}  \label{mf}
D_{\rm MF} =-a\Theta (-t)/b,\,\,\, \delta \widetilde{\Omega}_{\rm
MF }=- \frac{a^2}{2b}\Theta (-t),
\end{eqnarray}
where the step function $\Theta (-t)=1$ for $t<0$ and $\Theta
(-t)=0$ for $t>0$.

To be specific, discussing $T<T_c$ we further  consider the CFL
case. For the finite system of a large spherical size $R\gg \xi$,
with $\xi$ having the meaning of the coherence length, we obtain
\begin{eqnarray}\label{mf-sol}
d_{\alpha ,\rm MF}^i = \pm \frac{\delta_{\alpha}^i}{(N_c N_f)^{1/4}}
\sqrt{D_{\rm MF}}\,\Theta (-t)\, \mbox{th}
\left[\frac{(R-r)}{\sqrt{2}\,\xi}\right],\,\,\, \xi
=\sqrt{\frac{c}{|a|}},
\end{eqnarray}
where $N_c =3$ is the number of colors, $N_f =3$ is the number of
flavors and to be specific we assumed the simplest  structure 
$d_{\alpha ,\rm MF}^i \propto \delta_{\alpha}^i$.

Now we will consider  fluctuations. Below $T_c$ we present
$d^i_\alpha$, as $d^i_\alpha =d_{\alpha ,c}^i
+d^{i,\,\prime}_{\alpha}$, and above $T_c$, as
$d_{\alpha}^i=d^{i,\,\prime}_{\alpha}$, where index "c" labels the
classical solution. 


Since $\delta \widetilde{F} (V,T)=\delta
\widetilde{\Omega} (\mu , T),$ cf. \cite{LL}, the density of the
free energy  of the CFL phase expanded in
$(d^{i,\,\prime}_{\alpha })^2$ terms 
renders
\begin{eqnarray}\label{fr-F}
\delta \widetilde{F}&=& \delta \widetilde{F}_c +\delta
\widetilde{F}^{\,\prime}= \sum_{\alpha ,i}(-c_0|\partial_{\tau}
d_{\alpha ,\,c }^{i}|^2 +c |\nabla d_{\alpha ,\,c}^{i}|^2) +a D_c
+\frac{b_1 +b_2 /3}{2} D^2_c\nonumber
\\
&+&\frac{5}{3}(b_1 +b_2 /3)D_c N_c N_f \sum_{k}
|\phi_k^{\,\prime}|^2+N_c N_f \sum_{k} \left( -c_0 \omega^2
+c\vec{k}^{\,2} +a \right)|\phi_k^{\,\prime}|^2 \nonumber \\
&+&\nu_H \frac{b_1 +b_2 }{2}[N_c N_f \sum_{k}
|\phi_k^{\,\prime}|^2 ]^2 
.
\end{eqnarray}
The last term is introduced within the 
self-consistent Hartree approximation 
(within the $\Phi$ functional up to one vertex), $\nu_H =10/9$ accounts
different coefficients in $\Phi$ functional for the self-interaction
terms ($d^4$ - for the given field $d$ 
and $(d^i_{\alpha})^2 (d^j_{\beta})^2$ terms), cf
\cite{IKV}. 
We presented 
$d^{i,\prime}_{\alpha}=\sum_{k}d^{i,\,\prime}_{\alpha
,k}e^{-ik^\mu x_\mu}$, $k^\mu =(\omega , \vec{k})$, $x_\mu =({\tau},
-\vec{r})$, and introduced the notation
\begin{eqnarray}\label{phi-fl}
\sum_{i,\alpha ,k}|d_{\alpha ,k}^{i,\prime}|^2 =\sum_{k}
|\phi_k^{\,\prime}|^2 
=i\int \frac{d^4 k}{(2\pi)^4}\frac{1}{c_0
\omega^2 -c\vec{k}^{\,2} -m^2}.
\end{eqnarray}
In Matsubara technique the temperature dependence is introduced
by the replacements: $\omega \rightarrow \omega_n =2\pi i nT$,
$-i\int \frac{d^4
k}{(2\pi)^4}\rightarrow T\sum_{n=-\infty}^{n=\infty}\int \frac{d^3
k}{(2\pi)^3}$.
We restricted ourselves by the self-consistent Hartree approximation.
Effects of the
damping of fluctuations, being produced by the
two- and more vertex diagrams of $\Phi$, are beyond the scope
of this simple approximation. As we argue below, the main 
effects we discuss in
this paper are not essentially affected by the width terms.

Variation of eq. (\ref{fr-F}) over $\phi_k^{\,\prime}$
yields 
the spectrum of fluctuations
\begin{eqnarray}\label{eq-mfl}
c_0 \omega^2 -c\vec{k}^{\,2} -m^2 =0,
\end{eqnarray}
with the squared mass parameter
\begin{eqnarray}\label{mas} m^2 =\eta |a|=\frac{5}{3}(b_1 +b_2 /3
)D_c  +a+ \nu_H (b_1 +b_2 ) N_c N_f \sum_{k}
|\phi_k^{\,\prime}|^2 \,.
\end{eqnarray}
Notice that the real physical meaning of the
effective mass has the quantity
$m /\sqrt{c_0}$ rather than $m$, as follows from (\ref{eq-mfl}).

Solution of the equation for the fluctuating field presented in the
coordinate space is
characterized by the length scale $l=\xi/\sqrt{\eta }$ and by the
time scale $\widetilde{\tau} =\xi c_0^{1/2} (c\eta )^{-1/2}$.  Above the
critical point $D_c =0$, $ m^2 = a+O(\sum_{k}
|\phi_k^{\,\prime}|^2) >0$, and neglecting $|\phi_k^{\,\prime}|^2$ terms
one gets the parameter $\eta \simeq 1$.

Variation of (\ref{fr-F}) over $d^{i,*}_{\alpha ,c}$
yields the equation of motion for the classical field
\begin{eqnarray}\label{eq-mfcl}
&&-c_0
\partial^2_{\tau} d^{i}_{\alpha ,c} + c\Delta d^{i}_{\alpha ,c}
-\left(a+\frac{5}{3}(b_1 +b_2 /3)N_c N_f \sum_{k}
|\phi_k^{\,\prime}|^2\right)d^{i}_{\alpha ,c}\nonumber \\ 
&&-(b_1 +b_2 /3 )D_c
d^{i}_{\alpha ,c} =0.
\end{eqnarray}
Only dropping the term responsible for fluctuations we find $D_c
=D_{\rm MF}$, $d_{\alpha ,\,c}^i =d^i_{\alpha , \,\rm MF}$, $m^2
=m_{\rm MF}^2 =\eta_{\rm MF}|a|$ for $T<T_c$, cf. (\ref{mf}),
(\ref{mas}), and we obtain $\eta_{\rm MF} =2/3$.
In reality fluctuations affect the classical solution and
renormalize the critical temperature of the phase transition $T_c$
yielding $T_c^{\rm ren} < T_c$.
 More generally
\begin{eqnarray}\label{mren}
m^2 =m^2_{\rm MF}+\delta m^2,\,\,\,\, \delta m^2 =
-\frac{5}{3}(b_1 -b_2 /9)N_c N_f \sum_{k}
|\phi_k^{\,\prime}|^2 ,
\end{eqnarray}
for $T<T_c^{\rm ren}$. From (\ref{eq-mfcl}), (\ref{mren}) one finds
$D_c (m^2 =0)=0$, as the consequence of the self-consistency of our
approximation scheme.
As we have mentioned we neglected fluctuations of the gluon fields 
which change the nature of the phase transition but yield  only
a small jump of the order
parameter.

In order to demonstrate how fluctuations of the order parameter may
affect thermodynamic quantities let us calculate the contribution
of fluctuations to the specific heat density $\widetilde{C}_V
=-T\left(\frac{\partial^2 \delta \widetilde{F}} {(\partial
T)^2}\right)_V$, cf. \cite{LL}. The mean field contribution is
\begin{eqnarray}\label{mf-c}
\widetilde{C}_{V}^{\rm MF} =T\frac{a_0^2}{bT_c^{2}} \Theta (-t),
\end{eqnarray}
as it follows from (\ref{fe}), (\ref{fe-coef}) and (\ref{mf}).
Here and below we use an approximate expression $a\simeq a_0 t$
valid at $T$ near $T_c$, see  (\ref{fe-coef}). 
The 
contribution to the specific heat from the normal quark
excitations is suppressed as $\propto T^{-3/2}e^{-\Delta (0)/T}$ 
for $T\ll T_c$, cf. \cite{LP80}. We may reproduce correct low temperature
behavior multiplying
(\ref{mf-c})
by a form-factor 
$f\simeq 2e^{-\Delta (0)/T}(T_c/T)^{5/2}(1-\frac{2T}{3T_c})^{-1}$.

The fluctuation contribution $\widetilde{F}^{\,\prime}$ can be
determined with the help of the functional integration
\begin{eqnarray}\label{efV}
\mbox{exp}(-\delta W) =\int D \phi^{\,\prime} \,\mbox{exp}(-\delta
W[\phi^{\,\prime}]),
\end{eqnarray}
where $\delta W$ is an effective potential. Using (\ref{efV}),
(\ref{fr-F}) we obtain
\begin{eqnarray}\label{fl-f}
\delta \widetilde{F}^{\,\prime} =-iN_c N_f \int \frac{d^4
k}{(2\pi)^4}\mbox{ln}[c_0 \omega^2 -c\vec{k}^{\,2} -m^2].
\end{eqnarray}
Again within the Matsubara technique 
one still should do the replacement $\omega \rightarrow \omega_n =2\pi i nT$,
$-i\int \frac{d^4
k}{(2\pi)^4}\rightarrow T\sum_{n=-\infty}^{n=\infty}\int \frac{d^3
k}{(2\pi)^3}$.
We dropped an infinite constant term in (\ref{fl-f}). However expression
(\ref{fl-f}) still contains a divergent contribution. To remove
the regular term that does not depend on the closeness to
the critical point we  find the temperature derivative of
(\ref{fl-f}) (entropy per unit volume):
\begin{eqnarray}
\delta \widetilde{S}^{\,\prime}=\frac{\partial \delta
\widetilde{F}^{\,\prime}}{\partial T} = \frac{\partial
m^2}{\partial T} N_c N_f \sum_{k} |\phi_k^{\,\prime}|^2 .
\end{eqnarray}
The contribution of fluctuations to the specific heat density  is
then given by
\begin{eqnarray}\label{spec-gen}
 \widetilde{C}_V^{\,\prime}&=&
 -T\frac{\partial m^2}{\partial T} N_c N_f \frac{\partial \sum_{k}
|\phi_k^{\,\prime}|^2 }{\partial T}-T \frac{\partial^2
m^2}{\partial
T^2} N_c N_f \sum_{k} |\phi_k^{\,\prime}|^2 \nonumber \\
&\simeq& -T\frac{\partial m^2_{\rm MF}}{\partial T} N_c N_f
\frac{\partial \sum_{k} |\phi_k^{\,\prime}|^2 }{\partial T},\,\,\,
\frac{\partial m^2_{\rm MF}}{\partial T}\simeq -\eta_{\rm MF} a_0
/T_c .
\end{eqnarray}
In the second line (\ref{spec-gen}) we remained only the terms
quadratic in fluctuation fields, i.e. we assumed $|\frac{\partial
m^2_{\rm MF}}{\partial T}|\gg |\frac{\partial \delta m^2}{\partial
T}|$.

Now we are able to calculate the quantity $iG^{- -}(X=0)=\sum_{k}
|\phi_k^{\prime}|^2 $ and its temperature derivative, where $iG^{-
-}(X)$ is the time-ordered Green function. Using the
Matsubara replacement $\omega\rightarrow \omega_n =2\pi i nT$ and
the relation $\sum_{n}(y^2 +n^2)^{-1}=\frac{\pi}{y}\mbox{cth} (\pi
y)$, or the corresponding relation between the non-equilibrium
Green functions $iG^{- -}(X)$ and $\mbox{Im}G^{ret}$ at  finite
temperature, we arrive at the expression
\begin{eqnarray}\label{phiMat}
\sum_{k} |\phi_k^{\,\prime}|^2 =\frac{1}{2\sqrt{c_0}}\int
\frac{d^3 k}{(2\pi)^3}\frac{1}{\sqrt{c\vec{k}^{\,2}
+m^2}}\mbox{cth}\left( \frac{\sqrt{c\vec{k}^{\,2}
+m^2}}{2T\sqrt{c_0}}\right).
\end{eqnarray}
Using that $\mbox{cth}(y/2)=2n_B (y)+1$, where $n_B (y)=(e^y
-1)^{-1}$ are Bose occupations, and dropping the regular
contribution of quantum fluctuations we obtain
\begin{eqnarray}\label{phiMat-1}
\sum_{k} |\phi_k^{\,\prime}|^2_T =\frac{1}{\sqrt{c_0}}\int
\frac{d^3 k}{(2\pi)^3}\frac{1}{\sqrt{c\vec{k}^{\,2}
+m^2}}n_B\left( \frac{\sqrt{c\vec{k}^{\,2}
+m^2}}{T\sqrt{c_0}}\right).
\end{eqnarray}
By index $"T"$ we indicate the thermal contribution.

The integration is performed analytically in the limiting cases. Let
$m\gg T\sqrt{c_0}$. Then $n_B (y) \simeq e^{-y}$. In the very same
approximation one has $c\vec{k}^2 \ll m^2$ for typical momenta.
Then
\begin{eqnarray}\label{phi-it}
\sum_{k} |\phi_k^{\,\prime}|^2_T
=\frac{c_0^{1/4}}{8m}\left(\frac{2mT}{\pi
c}\right)^{3/2}\mbox{exp}\left(-
\frac{m}{T\sqrt{c_0}}\right),\,\,\,  m\gg 2T\sqrt{c_0}\,.
\end{eqnarray}
In the opposite limiting case, $m\ll 2T\sqrt{c_0}$, there are two
contributions to the integral (\ref{phiMat-1}), from the region of
typical momenta $c\vec{k}^{\,2} \sim m^2$ and from the region
$c\vec{k}^{\,2} \gg m^2$. They  can be easily separated, if one
calculates the auxiliary quantity $\frac{\partial \sum_{k}
|\phi_k^{\,\prime}|^2 }{\partial T}$. In the region
$c\vec{k}^{\,2} \sim m^2$ one may use an approximation  $n_B
(y)\simeq 1/y$ and
\begin{eqnarray}\label{phi-itht}
\frac{\partial \sum_{k} |\phi_k^{\,\prime}|^2 }{\partial
T}[c\vec{k}^{\,2} \sim m^2] \simeq -(4\pi c^{3/2})^{-1}
T\frac{\partial m}{\partial T}.
\end{eqnarray}
The region $c\vec{k}^{\,2} \gg m^2$ yields a regular term
\begin{eqnarray}\label{phi-itrel}
\sum_{k} |\phi_k^{\,\prime}|^2  [c\vec{k}^{\,2} \gg m^2]\simeq
\frac{T^2}{12}\sqrt{\frac{c_0}{c^3}},\,\,\,\,\,\, \frac{\partial
\sum_{k} |\phi_k^{\,\prime}|^2 }{\partial T} [c\vec{k}^{\,2} \gg
m^2]\simeq \frac{T}{6}\sqrt{\frac{c_0}{c^3}}\,.
\end{eqnarray}
General expression for the fluctuation contribution to the specific heat
is given by the  first line (\ref{spec-gen}) and can be resolved
with the help of the self-consistent solution of (\ref{mren}), 
(\ref{phiMat-1}) (or (\ref{phi-it}), (\ref{phi-itht}), (\ref{phi-itrel})
in the limiting cases). 
Assuming for rough estimates
that fluctuations can be described perturbatively and
putting $m^2 \simeq m^2_{\rm MF}=\eta_{\rm MF} |a|$, 
from the second line of (\ref{spec-gen})
we find for $T\ll \frac{m_{\rm
MF}}{\sqrt{c_0}}$:
\begin{eqnarray}\label{c-fllt}
\widetilde{C}^{\,\prime}_V \simeq \frac{N_c N_f \eta^{7/4}_{\rm
MF}\,\, c^{1/4}\,T^{3/2}}{4\,\pi^{3/2}\sqrt{2} \,c_0^{1/4}\,
|t|^{1/4}T_c^2} \left(\frac{a_0}{c}\right)^{7/4}\left(
1+\frac{2|t|T_c}{T}\right) \mbox{exp}\left(-\frac{\sqrt{\eta_{\rm
MF} a_0 |t|}}{T\sqrt{c_0}}\right)
\end{eqnarray}
and for $T\gg \frac{m_{\rm MF}}{\sqrt{c_0}}$:
\begin{eqnarray}\label{c-flht}
\widetilde{C}^{\,\prime}_V \simeq \frac{N_c N_f \eta^{3/2}_{\rm
MF}\,\, T^2}{8\pi T_c^2 |t|^{1/2}}
\left(\frac{a_0}{c}\right)^{3/2}+ \frac{N_c N_f  \eta_{\rm MF}
T^2}{6 T_c } \frac{a_0}{c}\left(\frac{c_0}{c}\right)^{1/2}\,.
\end{eqnarray}
In the weak coupling limit for the CFL phase
$\eta_{\rm MF} =2/3$, $a_0 /c =6\pi^2 T_c^2 /(7\zeta (3)/8)$, $c_0
=3c$. Eq.
(\ref{c-fllt}) holds in the low temperature limit $T\ll \pi
(|t|/3)^{1/2}\,T_c$, whereas eq. (\ref{c-flht}) is valid in the
opposite limit. The first term in (\ref{c-flht})
dominates over the second one 
for $|t| <0.7$, i.e., in the whole region of validity
of (\ref{c-flht}) the main term is the first one.
Its singular
behavior, $\sim 1/\sqrt{|t|}$, is typical for
the specific heat  in the vicinity of the critical point of the
second order phase transition, as in metallic superconductors. 

Also in the CFL phase ($T<T_c$) there is a contribution to the specific heat 
of Goldstone-like
excitations, cf. phonons in the ordinary condensed matter. 
For $T\gg m_{p.G}$, where
$m_{p.G}$ is the mass of the pseudo-Goldstone excitation,
we get
\begin{eqnarray}\label{c-G}
\widetilde{C}^{ p.G, \,\prime}_V =\frac{N_G 3^{3/2} \pi^2 T^3}{30},
\end{eqnarray} 
where $N_G$ is the number of pseudo-Goldstone modes.
One can see that the contribution (\ref{c-G}) is numerically small 
compared to those terms (cf. (\ref{c-fllt}), (\ref{c-flht})) we have
evaluated for temperatures within the fluctuation
region. 

Comparing the mean field (\ref{mf-c}) and the fluctuation
(\ref{spec-gen}) contributions to the specific heat (in the low
and high temperature  limiting cases  one may use eqs.
(\ref{c-fllt}), (\ref{c-flht})) we may estimate the 
fluctuation temperature $T_{\rm fl,<}^{C}<T_c$, at which the
contribution of fluctuations of the order parameter becomes to be
as important as the mean field one (so called 
Ginzburg - Levanyuk criterion),
\begin{eqnarray}\label{fl-mean}
\widetilde{C}^{\,\prime}_V \simeq \widetilde{C}_V^{\rm MF},\,\,\,
{\rm for}\,\,\, T<T_c .
\end{eqnarray}
Fluctuations dominate for $T>T_{\rm fl,<}^{C}$. For typical values
$\mu_q \sim (350\div 500)$~MeV and for $T_c \gsim (50\div 70)$~MeV
in the weak coupling limit from
(\ref{fl-mean}), (\ref{c-fllt}) we estimate $T_{\rm fl,<}^{C}\simeq (0.6\div
0.8)T_c$.  If we took into account the suppression factor $f$
of the mean field term $\propto e^{-\Delta (0)/T}$, a decrease
of the mass $m$ due to the fluctuation contribution  (cf. (\ref{mren})),
and the pseudo-Goldstone contribution (\ref{c-G}),
we would get
still smaller value of $T_{\rm fl,<}^{C}$ ($\lsim 0.5T_c$).
We see that fluctuations start to contribute at
temperatures when one can still use approximate expressions
(\ref{c-fllt}), (\ref{phi-it}) 
valid in the low temperature limit. Thus the time
(frequency) dependence of the fluctuating fields is important in
case of CSC.

In the condensed matter physics  one usually performs calculations
in the high temperature limit. In this limit one neglects the time
(frequency) dependent terms considering quasi-static 
thermal fluctuations of the order
parameter. Then the fluctuation
contribution is determined with the help of the functional
integration
\begin{eqnarray}\label{fl-o}
\mbox{exp}(-\delta F^{\,\prime}/T) &=&\int D d^{\,\prime}
\,\mbox{exp}(-\delta F^{\,\prime}[d^{\,\prime}]/T),\,\,\,\, \\
\delta
F^{\,\prime}[d^{\,\prime}]&=& \sum_{i, \alpha ,
\vec{k}}(c\vec{k}^{\,2} +\eta |a|)(d_{\alpha
,\,\vec{k}}^{i,\,\prime} )^2 .\nonumber
\end{eqnarray}
The integration yields
\begin{eqnarray}\label{fl-o1} \delta
\widetilde{F}^{\,\prime}= TN_f N_c\int \frac{d^3 k }{(2\pi
)^3}\mbox{ln}[c\vec{k}^{\,2} +\eta |a|].
\end{eqnarray}
Comparison of eqs. (\ref{fl-o1}) and  (\ref{fl-f}) shows that the
high temperature expression (\ref{fl-o1}) is recovered, if
one drops all the terms except $n=0$ in the corresponding
Matsubara sum over $\omega_n =2\pi i nT$ in eq. (\ref{fl-f}). The
contribution of $n\neq 0$ terms to the $\sum_{k}
|\phi_k^{\,\prime}|^2$ is suppressed in the limit $c_0 4\pi^2 T^2
\gg m^2$. Then one immediately arrives at the specific heat given
by the first term of (\ref{c-flht}). Thus in our case one may
suppress the frequency dependence of fluctuations only for $|t|\ll
1$.

Simplifying, the energy width of the fluctuation region, where the
fluctuation effects prevail over the mean field ones, is usually
estimated following the  Ginzburg criterion. The probability of
the fluctuation in the volume $V_{\rm fl}$ is given by $W\sim
\mbox{exp}(-\delta \Omega (V_{\rm fl}) /T)$. It is $\sim 1$
for $\delta \Omega (V_{\rm fl}) \sim T$, where $\delta \Omega
(V_{\rm fl})$ (the contribution to the thermodynamic potential in
the corresponding variables) is the work necessary to prepare the
fluctuation within the volume $V_{\rm fl}$. The minimal size of
the fluctuation region characterized by an order
parameter $d \sim \sqrt{D_{\rm MF}}$ is $\xi(T)\sqrt{2/\eta}$, cf.
(\ref{mf-sol}), (\ref{eq-mfl}). Thus, taking $\delta \Omega
(V_{\rm fl})=\delta \Omega_{\rm MF} (V_{\rm fl})= T_{\rm fl}^{\rm
G}$ for the typical temperature $T_{\rm fl}^{\rm G}$, when
fluctuations start to dominate, we obtain
\begin{eqnarray}\label{Ginz}
T_{\rm fl}^{\rm G}\simeq \frac{a^2}{2b}\frac{4\pi (\sqrt{2/\eta}\,
\xi (T_{\rm fl}^{\rm G}))^3}{3}\simeq \frac{ 4\pi a_0^{1/2}c^{3/2}
|t(T_{\rm fl}^{\rm G})|^{1/2}}{\eta^{3/2}b} \frac{\sqrt{2}}{3}.
\end{eqnarray}
Although the above estimate is very rough we took care of all
the numerical factors. This allows us to notice that the  value
$T_{{\rm fl},<}^{C}$ estimated from (\ref{fl-mean}), if one uses
eq. (\ref{c-flht}) for $\widetilde{C}^{\,\prime}_V $ remaining
there only the first term, is $T_{{\rm fl},<}^{C}\simeq 0.5
|t(T_{{\rm fl},<}^{C})/t(T_{\rm fl}^{\rm G})|^{1/2} T_{\rm
fl}^{\rm G}$ for $N_c =N_f =3$. All the dependencies on the
parameters in expressions for $T_{{\rm fl},<}^{C}$ and
$T_{\rm fl}^{\rm G}$ were proven to be essentially the same.
Fluctuation region is determined by the Ginzburg-Levanyuk
parameter $Gi =|T_c -T_{\rm fl}^{\rm C}|/T_c$. 
The larger the value $Gi$ is the wider is the
fluctuation region. For clean conventional superconductors
\cite{LV} $Gi =10^{-12}\div
10^{-14}$, whereas for  superfluid He$^4$ and in our case $Gi \sim 1$.

Now we are able to discuss the validity of the self-consistent Hartree
approximation in our problem. Adding the contribution to $\Phi$ functional with
two vertices produces the sun-set diagram in the di-quark self-energy.
In such a way beyond the Hartree approximation
there appears the width term in the self-energy (behaving as 
$-i\gamma \omega$ for small $\omega$). Such a 
term governs the slow relaxation of the 
order parameter in the  phase transition phenomena outside the equilibrium.
In our case (thermal equilibrium)
this term can be dropped compared to the $c_0\omega^2$ term,
which we have in the thermodynamic potential (\ref{fe}) from the very
beginning,
at least in both the low and high temperature limits. 
Indeed, in the low temperature limit three Green functions entering the
sun-set diagram produce an extra exponentially small particle occupation
factor compared to that governs the Hartree term (\ref{phi-it}).
Near $T_c$, i.e. in the high temperature limit, 
we may remain only $n=0$ term in
the Matsubara sum over frequencies, as we have argued,  
thus suppressing both the
linear and the quadratic  terms in $\omega$.

All above expressions
for fluctuating quantities (except the vanishing of the
pseudo-Goldstone contribution above $T_c$ and the appearance of an extra
diquark decay contribution to the width) are also
valid for $T>T_c$, if one puts $D_c =0$ in general expressions
or $m^2_{\rm MF} =a>0$, $D_c =0$,  $\eta
=\eta_{\rm MF}=1$ in the corresponding approximate 
expressions.
Above $T_c$ we
should compare the fluctuation contribution to a thermodynamic
quantity with the quark and gluon contributions of the normal
state of the quark-gluon plasma. In the weak coupling limit for
the specific heat density of the quark-gluon plasma one has, cf.
\cite{Mul},
\begin{eqnarray}\label{norm-c}
\widetilde{C}^{qg}_V \simeq 6 \mu_q^2 T  +\frac{42\pi^2
T^3}{15}+\frac{32\pi^2 T^3}{15}.
\end{eqnarray}
The first two terms are quark contributions and the third term is the
gluon contribution. We omitted a contribution of strange quarks
which is rather small for $T<m_s$ and we neglected the temperature
dependent effective gluon mass. We also disregarded the $\alpha_s$
corrections to the quark and gluon terms since such corrections
were not taken into account for condensate quantities.

Above $T_c$, eqs. (\ref{c-fllt}) and (\ref{c-flht}) yield rather
smooth functions of $T$ except the region $t\ll 1$. The fluctuation
region is rather wide since $\widetilde{C}^{\,\prime}_V \sim
\widetilde{C}^{qg}_V$ even at $T$ essentially larger then  $T_c$.
The appearance of an extra channel of the di-quark 
decay width, $2\gamma_{\rm dec}\omega$,
beyond the Hartree approximation
does not qualitatively change the situation since $\gamma_{\rm dec}$
is a regular function of $T$,
$\gamma_{\rm dec}\propto T-T_c$, and vanishes in the critical point. Thus
the applicability of the high temperature limit 
(one can put $\omega_n =0$ in the
Matsubara sum) is preserved in a wide temperature region, $T\sim T_c$, 
\cite{LV}. 
Only far above $T_c$ situation might be changed.
The main uncertainty comes from the values of  coefficients (\ref{fe-coef}) 
which were derived for
$T$ rather 
near  $T_c$ and in the weak coupling limit. Bearing all this in mind 
we estimate the value of the fluctuation temperature
$T_{{\rm fl},>}^C \sim 2T_c$ for typical value $T_c /\mu_q \sim
0.1\div 0.3$. The higher the value $T_c  $ is, the stronger is the
contribution of fluctuations at the given ratio $T/T_c$.

Thus we see that {\em fluctuations of the di-quark gap may essentially
contribute to the thermodynamic quantities even well below $T_c$
and  above  $T_c$.}

Note that several simplifying  assumptions were done. The
coefficients
(\ref{fe-coef}) were derived for $|t|\ll 1$ in the weak coupling
limit  neglecting the strange quark mass, but applied 
in a wider temperature region for may be not sufficiently large $\mu_q$.  
Only Gaussian
fluctuations were taken into account within the self-consistent 
Hartree approximation.  Thus,  di-quark width effects were assumed to be
suppressed. 
We used the expansion of the
thermodynamic potential up to the fourth order terms in the mean
field neglecting a small jump in the order parameter 
due to gluon fluctuations. We incorporated in the thermodynamic quantities
only the terms which have a tendency to 
an irregular
behavior near $T_c$, whereas above $T_c$ the short range 
correlations begin to be more and more important with the increase of the
temperature. 
Therefore one certainly should be cautious applying
above rough estimates outside the region of their quantitative
validity. However, as we know from the experience of the
condensate matter physics, see \cite{LV}, such extrapolation equations work
usually not too bad even for temperatures well below and above
$T_c$.

So far we have discussed the specific behavior of fluctuations of
the order parameter at fixed temperature and density.
There are also fluctuations of the temperature and the local quark
density.  They are statistically independent quantities \cite{LL},
$<\delta T\delta\rho_q
>=0$, and their mean squares are
\begin{eqnarray}\label{fl-tr}
<(\delta T)^2 > =\frac{T^2}{V_{\rm fl} \widetilde{C}_V },\,\,\,
<(\delta \rho_q)^2 > =\frac{T\rho_q}{V_{\rm
fl}}\left(\frac{\partial \rho_q} {\partial P}\right)_T.
\end{eqnarray}
The averaging is done over the volume, $P$ is the pressure,
$V_{\rm fl}$ is as above the volume related to the fluctuation. In
the limiting case $\widetilde{C}^{\,\prime}_V \ll
\widetilde{C}^{qg}_V$, we obtain $<(\delta T)^2 >/T^2 \simeq
\rho_q /(N^{\rm fl}_q \widetilde{C}_V^{qg})$, where $N^{\rm fl}_q$
is the number of quarks involved in the volume $V_{\rm fl}$. Thus,
far from the critical point the contribution of fluctuations is
suppressed as $\sqrt{<(\delta T)^2
>/T^2}\propto 1/\sqrt{N^{\rm fl}_q}$. This standard fluctuation
behavior is essentially changed for $\widetilde{C}^{\,\prime}_V
\gsim \widetilde{C}^{qg}_V$. Then we get even larger suppression
of the temperature fluctuations. For $|t|\ll 1$ in the high
temperature limit we obtain $\sqrt{<(\delta T)^2
>/T_c^2} \sim |t|\rightarrow 0$ for
$|t|\rightarrow 0$, where we assumed that fluctuations are most
probable within the typical fluctuation volume $V_{\rm
fl}=4\pi \xi^3 /3$ and $\xi \propto 1/\sqrt{|t|}$ for $t
\rightarrow 0$. Thus, if the system is at the temperature $T$ in a
narrow vicinity of $T_c$, fluctuations of the temperature are
significantly suppressed. Being 
formed at $0< -t\ll 1$ the
condensate region evolves  very slowly, since the heat transport is then
delayed (the typical evolution time is roughly $\tau\propto
1/\sqrt{|t|}$). At temperatures $T$ outside a narrow
vicinity of $T_c$, fluctuations of the temperature resulting in
a significant decrease of the temperature in the volume $V_{\rm fl}$  
are rather probable, $\sqrt{<(\delta T)^2
>/T_c^2} \sim 1$. This is the
consequence of a very short coherence length and, thus, not too
large number of quarks contained in the volume $V_{\rm fl}$, $\xi
\simeq 0.13/(T_c \sqrt{|t|})$ being $\simeq 0.5$~fm for $T_c
\simeq 50$~MeV and for $|t|\sim 1$.  
Fluctuations of the quark density are also large
for typical $|t|\sim 1$, due to the smallness of $V_{\rm fl}$.
Thereby we  argue that the system may produce
the di-quark condensate regions of the typical size $\xi$,  thus
feeling the possibility of the phase transition even if its
temperature and density are in average rather far  from the
critical values.  

Anomalous behavior of fluctuations might
manifest itself in the event-by-event analysis of the heavy ion
collision data. In small ($l$) size
systems, $l<\xi$, (zero dimension case would be $l\ll \xi$)
the contribution of fluctuations of the order parameter
to the specific heat is still increased, see
\cite{LV}.
The anomalous behavior of the specific heat may affect the heat
transport. If thermalization happened at an initial heavy ion
collision stage, 
and a large size system
expands rather slowly, its evolution is governed by the
approximately constant value of the entropy. Due to a large
contribution to the specific heat (and to the entropy) of the
di-quark fluctuations, an  extra decrease of the temperature may
occur  
resulting in an essential slowing of the fireball  expansion
process (due to smaller pressure). 
Having the di-quark quantum numbers, fluctuations of
the gap  may affect the di-lepton production rate from the
quark-gluon plasma, as it has been noted in \cite{KKKN02}.

We would like once more
to emphasize that the coefficients of the thermodynamic
potential (\ref{fe}) obtained in the weak
coupling limit might be essentially modified, in case if we applied the
results to the fireball produced
in a heavy ion collision, somehow changing our conclusions. E.g.,
the quark chemical potential decreases with the temperature. It results in an
additional increase of the contribution of fluctuations, cf. dependences 
of the coefficients (\ref{fe-coef}) on
$\mu_q$. Thus we may need to analyze
the  strong coupling limit instead of the weak coupling limit we discussed. 
If $\mu_q$ becomes
smaller than the strange quark mass we come from the possible 3SC phases 
to the 2SC phases.
A more general discussion of modifications
beyond the framework of the weak coupling limit can be
found in \cite{P00}. A discussion how the
coefficients (\ref{fe}) may in general vary within the 
Ginzburg - Landau approach is given in \cite{IB}.      
Our above arguments are admittedly speculative, and mean only to demonstrate a
possibility of their application to heavy ion collisions and to an initial
stage evolution of the neutron star.
A more detailed treatment of the problem definitely needs a further work.

Other quantities associated with second derivatives of the
thermodynamic potential are also enhanced near the critical point
demonstrating typical $1/\sqrt{|t|}$ behavior, cf. \cite{LP80}. 
However numerical
coefficients depend strongly on what quantity is
studied.
E.g. fluctuation contributions above $T_c$ to the
color diamagnetic susceptibilities
\begin{eqnarray}\label{fl-g}
\chi_{\alpha} =-\left( \partial^2 \delta F /\partial
{\vec{\cal{H}}}_{\alpha}^2 \right)_{{\vec{\cal{H}}}_{\alpha} =0},\,\,\,
{\vec{\cal{ H}}}_{\alpha} =\mbox{curl}{\vec{\cal{A}}}_{\alpha}\,,
\end{eqnarray}
are proven to be  $ \ll 1$ everywhere except very narrow vicinity
of the critical point. In spite of a smallness, as we know, 
in ordinary clean superconductors the
fluctuation diamagnetism turns out to be of the order of the Pauli
paramagnetism even far from the transition \cite{LV}.

Concluding, within  the Ginzburg - Landau approach we estimated
the fluctuation energy region at the CSC phase transition. The quantitative
estimates are based on the values of
parameters derived in the weak coupling limit for 
fluctuations of the CFL order parameter. Qualitative results survive
also for fluctuations of other possible phases.
We found that 
the frequency
dependence of fluctuations is important for CSC in a wide
temperature region. Fluctuations may contribute essentially to the
specific heat even at $T$ rather far below and above $T_c$. We
estimated $T_{\rm fl, <} \lsim 0.5 T_c$ and $T_{\rm fl,
>} \sim 2 T_c$. Our rough estimates show that the high temperature CSC could
manifest itself through fluctuations of di-quark gap in the course
of the heavy ion collisions at SIS200, if the critical temperature
of the phase transition is indeed rather high, $T_c \gsim (50\div
70)$~MeV. However quantitative results depend on the values of
several not sufficiently known parameters and  further studies are still
needed to arrive at  definite conclusions.


The author thanks T. Kunihiro and M.F.M. Lutz
for the discussions. He acknowledges
the hospitality and support of GSI Darmstadt. The work has been
supported in part by DFG (project 436 Rus 113/558/0-2), and by
RFBR  grant NNIO-03-02-04008.



\end{document}